\documentclass[10pt,journal,letterpaper,compsoc]{IEEEtran}
\IEEEoverridecommandlockouts

\usepackage{amsmath}
\usepackage{amssymb}
\usepackage{amsfonts}
\usepackage[dvips]{graphicx}
\usepackage{subfig}
\usepackage{algorithm}
\usepackage{algorithmic}
\usepackage{slashbox}
\usepackage{tabularx}
\usepackage{booktabs}
\usepackage{multirow}
\usepackage[nocompress]{cite}
\usepackage{color}
\newtheorem{prop}{\textbf{Proposition}}[section]

\newtheorem{cor}{Corollary}

\newtheorem{lm}{Lemma}[section]

\newtheorem{thm}{\textbf{Theorem}}

\newcommand{\bthm}{\begin{thm}}
\newcommand{\ethm}{\end{thm}}

\newcommand{\bcor}{\begin{cor}}
\newcommand{\ecor}{\end{cor}}
\newcommand{\bprop}{\begin{prop}}
\newcommand{\eprop}{\end{prop}}
\newcommand{\blm}{\begin{lm}}
\newcommand{\elm}{\end{lm}}
\newcommand{\beq}{\begin{equation}}
\newcommand{\eeq}{\end{equation}}
\newcommand{\ber}{\begin{eqnarray}}
\newcommand{\eer}{\end{eqnarray}}

\newenvironment{proof1}{\begin{trivlist}\item[]{\bf Proof:\hspace{2mm}}}{\hfill$\blackbox$\end{trivlist}}

\newcommand{\blackbox}{\vrule height7pt width5pt depth1pt}

\newcommand{\bit}{\begin{itemize}}
\newcommand{\eit}{\end{itemize}}
\newcommand{\ben}{\begin{enumerate}}
\newcommand{\een}{\end{enumerate}}
\newcommand{\bdesc}{\begin{description}}
\newcommand{\edesc}{\end{description}}
\newcommand{\beqarrn}{\begin{eqnarray*}}
\newcommand{\eeqarrn}{\end{eqnarray*}}
\newenvironment{proofof}[1]{\begin{trivlist}\item[]{\bf Proof of #1:\hspace{2mm}
}}{\hfill\blackbox\end{trivlist}}
\newcommand{\bproofof}{\begin{proofof}}
\newcommand{\eproofof}{\end{proofof}}
\newenvironment{rem}{\begin{trivlist}\item[]{\bf
Remark:}\hspace{4mm}}{\end{trivlist}}
\newcommand{\brem}{\begin{rem}}
\newcommand{\erem}{\end{rem}}
\newenvironment{rems}{\begin{trivlist}\item[]{\bf
Remarks}\begin{itemize}}{\end{itemize}\end{trivlist}}
\newcommand{\brems}{\begin{rems}}
\newcommand{\erems}{\end{rems}}
\newtheorem{fact}{Fact}
\newcommand{\bfact}{\begin{fact}}
\newcommand{\efact}{\end{fact}}
\newtheorem{examp}{Example}
\newcommand{\bexamp}{\begin{examp}\rm}
\newcommand{\eexamp}{\end{examp}}
\newtheorem{defn}{Definition}
\newcommand{\bdefn}{\begin{defn}\rm}
\newcommand{\edefn}{\end{defn}}

\newtheorem{prob}{Problem}
\newcommand{\bprob}{\begin{prob}}
\newcommand{\eprob}{\end{prob}}

\newcommand{\bvtm}{\begin{verbatim}}
\newcommand{\bfig}{\begin{figure}}
\newcommand{\efig}{\end{figure}}
\newcommand{\bcen}{\begin{center}}
\newcommand{\ecen}{\end{center}}

\long\def\comment#1{}

\begin{document}

\title{{Distributed Stochastic Power Control in Ad-hoc Networks: A Nonconvex Case}}


\author {Lei Yang, 
         Yalin E. Sagduyu, 
         Junshan Zhang, 
         and Jason H. Li 
\IEEEcompsocitemizethanks{
\IEEEcompsocthanksitem Lei Yang and Junshan Zhang are with the School of ECEE, Arizona State University, Tempe, AZ, 85287, USA (e-mail: lyang55@asu.edu; Junshan.Zhang@asu.edu).
\IEEEcompsocthanksitem Yalin E. Sagduyu and Jason H. Li are with Intelligent Automation, Inc., Rockville, MD 20855, USA (e-mail: ysagduyu@i-a-i.com; jli@i-a-i.com).
\IEEEcompsocthanksitem Part of this paper will be presented at the IEEE International Conference on Communications, ICC 2011 \cite{yang:2011}.
}
}


\IEEEcompsoctitleabstractindextext{%
\begin{abstract}
Utility-based power allocation in wireless ad-hoc networks is inherently nonconvex because of the global coupling induced by the co-channel interference. To tackle this challenge, we first show that the globally optimal point lies on the boundary of the feasible region, which is utilized as a basis to transform the utility maximization problem into an equivalent max-min problem with more structure. By using extended duality theory, penalty multipliers are introduced for penalizing the constraint violations, and the minimum weighted utility maximization problem is then decomposed into subproblems for individual users to devise a distributed stochastic power control algorithm, where each user stochastically adjusts its target utility to improve the total utility by simulated annealing. The proposed distributed power control algorithm can guarantee global optimality at the cost of slow convergence due to simulated annealing involved in the global optimization. The geometric cooling scheme and suitable penalty parameters are used to improve the convergence rate. Next, by integrating the stochastic power control approach with the back-pressure algorithm, we develop a joint scheduling and power allocation policy to stabilize the queueing systems. Finally, we generalize the above distributed power control algorithms to multicast communications, and show their global optimality for multicast traffic.
\end{abstract}
\begin{keywords}
Distributed Power Control, Nonconvex Optimization, Extended Duality Theory, Simulated Annealing, Queue Stability, Unicast Communications, Multicast Communications.
\end{keywords}
}
\maketitle

\section{Introduction} \label{sec:introduction}
\IEEEPARstart{T}{he} broadcast nature of wireless transmissions makes wireless networks susceptible to interference, which deteriorates quality of service (QoS) provisioning. Power control is considered as a promising technique to mitigate interference. One primary objective of power control is to maximize the system utility that can achieve a variety of fairness objectives among users\cite{chiang:2008,julian:2002,chiang:2007,xiao:2003}.
However, maximizing the system utility, under the physical interference model, often involves nonconvex optimization and it is known to be NP-hard, due to the complicated coupling among users through interference \cite{luo:2008}.

Due to the nonconvex nature of the power control problem, it is challenging to find the globally optimal power allocation in a distributed manner. Notably, \cite{hande:2008,huang:2006} devised distributed power control algorithms to find power allocations that satisfy the local optimality conditions, but global optimality could not be guaranteed in general, except for some special convexifiable  cases (e.g., with strictly increasing log-concave utility functions). Another thread of work applied game-theoretic approaches to power control by treating it as a non-cooperative game among transmitters \cite{saraydar:2002,alpcan:2002}. However, distributed solutions that converge to a Nash equilibrium may be suboptimal in terms of maximizing the total system utility. Different from these approaches, \cite{qian:2009} proposed a globally optimal power control scheme, named MAPEL, by exploiting the monotonic nature of the optimization problem. However, the complexity and the centralized nature of MAPEL hinder its applicability in practical scenarios, and thus it can be treated rather as a benchmark for performance evaluation in distributed networks.

To find the globally optimal power allocation in a distributed setting, an interesting work \cite{qian:2010} has proposed the SEER algorithm based on Gibbs sampling \cite{geman:1984}, which can approach the globally optimal solution in an asymptotic sense when the control parameter in Gibbs sampling tends to infinity. Notably, for each iteration in the SEER algorithm, each user utilizes Gibbs sampling to compute its transition probability distribution for updating its transmission power, where the requirement for message passing and computing the transition probability distribution in each iteration can be demanding when applied to ad-hoc communications.

A challenging task in  distributed power control in ad-hoc networks is to reduce the amount of message passing while preserving the global optimality. In this paper, we tackle this challenge by combining recent advances in extended duality theory (EDT) \cite{chen:2008} with simulated annealing (SA) \cite{Kirkpatrick:1983}. Compared with the classical duality theory with nonzero duality gap for nonconvex optimization problems, EDT can guarantee zero duality gap between the primal and dual problems by utilizing nonlinear Lagrangian functions. This property allows for solving the nonconvex problem by its \emph{extended dual} while preserving the global optimality with distributed implementation. Furthermore, as will be shown in Section II, for the subproblem of each individual user, the extended dual can then be solved through stochastic search using SA. In particular, we first transform the original utility maximization problem into an equivalent max-min problem. This step is based on the key observation that in the case with continuous and strictly increasing utility functions, the globally optimal solution is always on the boundary of the feasible (utility) region. Then, appealing to EDT and SA, we develop a distributed stochastic power control (DSPC) algorithm that stochastically searches for the optimal power allocation in the neighborhood of the feasible region's boundary, instead of bouncing around in the entire feasible region.

Specifically, we first show that DSPC can achieve the global optimality in the underlying nonconvex problem, although the convergence rate can be slow (but this is clearly due to the slow convergence nature of SA). Then, to improve the convergence rate of DSPC, we propose an enhanced DSPC (EDSPC) algorithm that employs the geometric cooling schedule and performs a careful selection of penalty parameters. As a benchmark for performance evaluation, we also develop a centralized algorithm to search for the globally optimal solution over simplices that cover the utility region. The performance gain is further verified by comparing our distributed algorithms with MAPEL \cite{qian:2009}, SEER \cite{qian:2010}, and ADP \cite{huang:2006} algorithms. Worth noting is that the proposed DSPC and EDSPC algorithms do not require any knowledge of channel gains, which is typically needed in existing algorithms, and instead they need only the Signal-to-Interference-plus-Noise (SINR) feedback for adaptation.

Next, we integrate the above distributed power control  with the back-pressure algorithm \cite{ephremides:1992} and devise a joint scheduling and power allocation policy for improving the stability in the presence of dynamic packet arrivals and departures. This policy fits into the dynamic back-pressure and resource allocation framework and enables distributed utility maximization without extra technical conditions \cite{neely:2005} \cite{modiano:2011}.
Then, we generalize the study to  consider multicast communications, where a single transmission may simultaneously deliver packets to multiple recipients \cite{wieselthier:2000}. Specifically, we extend DSPC and EDSPC algorithms to multicast communications with distributed implementation, and show that these algorithms can also achieve the global optimality in terms of jointly maximizing the minimum rates on bottleneck links in different multicast groups.

The rest of the paper is organized as follows. In Section \ref{sec:pc_unicast}, we first introduce the system model, establish the equivalence between the utility maximization problem and its max-min form, and then develop both centralized and distributed algorithms for the max-min problem. Next, in Section \ref{sec:stability}, building on these power control algorithms, we develop a joint scheduling and power allocation policy to stabilize queueing systems. The generalization to multicast communications is presented in Section \ref{sec:multicast}. We conclude the paper in Section \ref{sec:conclusion}.

\section{Power Control For Unicast Communications}\label{sec:pc_unicast}
\subsection{System Model}

\begin{figure}[t]
\begin{center}
\vspace{1.8cm}\hspace{3.0cm} {\includegraphics[scale=0.6]{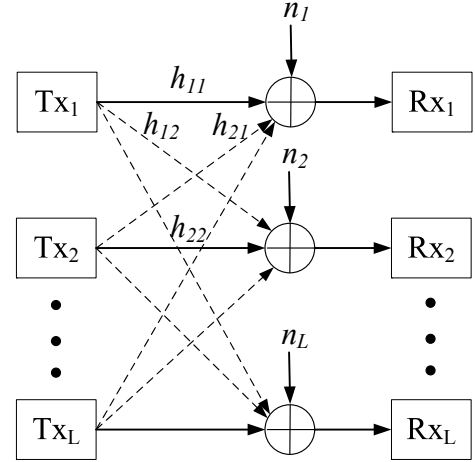}}\hspace{-0.0cm}
\vspace{5.5cm} \caption{System model.}\vspace{0cm}
\label{fig:model}
\end{center}
\end{figure}

We consider an ad-hoc wireless network with a set $\mathcal{L}=\{1,...,L\}$ of links, where the channel is interference-limited, and all $L$ links treat interference as noise, as illustrated in Fig. \ref{fig:model}. Such a model of communication is also applicable to cellular networks \cite{chiang:2008}. Each link consists of a dedicated transmitter-receiver pair.\footnote{We use the terms ``user'' and ``link'' interchangeably throughout.} We denote by $h_{lk}$ the fixed channel gain between user $l$'s transmitter and user $k$'s receiver, and by $p_l$ the transmission power of link $l$ with $P_l^{\max}$ being its maximum power constraint. For static channels, the received SINR for the $l$th user with a matched filter receiver is given by
\begin{equation}
\gamma_l(\mathbf{p})=\frac{h_{ll}p_l}{n_l+\sum_{k\ne l}{h_{kl}p_k}},
\end{equation}
where $\mathbf{p}=(p_1,...,p_L)$ is a vector of the users' transmission powers and $n_l$ is the noise power. Accordingly, the $l$th user receives the utility $U_l(\gamma_l)$, where $U_l(\cdot)$ is continuous and strictly increasing. We assume that each user $l$'s utility is zero when $\gamma_l=0$, i.e., $U_l(0)=0$. For ease of reference, some notation is listed in Table \ref{tab:notations}.\footnote{We use bold symbols (e.g., $\mathbf{p}$) to denote vectors and calligraphic symbols (e.g., $\mathcal{L}$) to denote sets.}

\begin{table}
\centering
\begin{tabular}{|c|c|}
\hline
\textbf{Notation} & \textbf{Definition}\\
\hline
$\mathcal{L}$ & set of links\\
\hline
$L$ & total number of links \\
\hline
\multirow{2}{*}{$h_{lk}$} & channel gain from link $l$'s transmitter\\
& to link $k$'s receiver\\
\hline
$\mathbf{H}$ & link gain matrix\\
\hline
$p_l$ (in vector $\mathbf{p}$) & transmission power of link $l$\\
\hline
$n_l$ (in vector $\mathbf{n}$) & noise power for link $l$\\
\hline
$\gamma_l$ & SINR of link $l$\\
\hline
$\gamma_l(\cdot)$ & SINR function of link $l$\\
\hline
$U_l(\cdot)$ & utility function of link $l$\\
\hline
$x_l$ (in vector $\mathbf{x}$) & ratio of link $l$'s utility to the total network utility\\
\hline
$r_l$ (in vector $\mathbf{r}$) & transmission rate of link $l$\\
\hline
$r_l(\cdot)$ & transmission rate function of link $l$\\
\hline
$\alpha$, $\beta$ & penalty multipliers\\
\hline
\end{tabular}
\caption{Summary of the notations and definitions.}\label{tab:notations}
\end{table}

\subsection{Network Utility Maximization}
We seek to find the optimal power allocation $\mathbf{p}^*$ that maximizes the overall system utility subject to the individual power constraints, given by the following optimization problem:
\begin{equation}%
\begin{array}
[c]{lll}%
&{\text{maximize} }&\sum_{l\in\mathcal{L}}U_l(\gamma_l(\mathbf{p}))
\\
&\text{subject to}& 0\leq p_l\leq P_l^{\max}, \forall l\in\mathcal{L}
\\
& \text{variables}& \{\mathbf{p}\}.
\end{array}
\label{eq:sum}
\end{equation}

In general, (\ref{eq:sum}) is a nonconvex problem\footnote{For some special utility functions $U_l(.)$, (\ref{eq:sum}) can be transformed into a convex problem \cite{chiang:2007}. In this paper, we focus on the nonconvex case that cannot be transformed to a convex problem by change of variables.}. In particular, if the utility function is the Shannon rate achievable over Gaussian flat fading channels, namely  $U_l(\gamma_l(\mathbf{p}))=w_l\log(1+\gamma_l(\mathbf{p}))$, where $w_l>0$ is a weight associated with user $l$, (\ref{eq:sum}) boils down to the weighted sum rate maximization problem, which is known to be nonconvex and NP-hard \cite{luo:2008}. Note that the weights can serve as the fairness measures\cite{mo:2000} for different scenarios. In particular, in queueing systems, for arrival rates within the stability region, packet queues can be stabilized by solving this weighted sum rate maximization problem, when the instantaneous queue lengths are chosen as the weights. In Section \ref{sec:stability}, we will discuss how to stabilize the packet queues by integrating our distributed power control algorithms with the back-pressure algorithm.

Let $\mathcal{F}$ denote the feasible utility region, where for each point $\mathbf{U}=(U_1,...,U_L)$ in $\mathcal{F}$, there exists a power vector $\mathbf{p}$ such that $U_l=U_l(\gamma_l(\mathbf{p}))$ for all $l\in\mathcal{L}$. The feasible utility region $\mathcal{F}$ is nonconvex, and in general, finding the globally optimal solution to (\ref{eq:sum}) in $\mathcal{F}$ is challenging. In the following example, we illustrate the geometry of $\mathcal{F}$ for the utility $U_l(\gamma_l(\mathbf{p}))=w_l\log(1+\gamma_l(\mathbf{p}))$ and evaluate the solutions to (\ref{eq:sum}) given by some existing power control approaches discussed in Section \ref{sec:introduction}.

\begin{figure}[t]
\centering
\hspace{-0cm}
 {\includegraphics[scale=0.5]{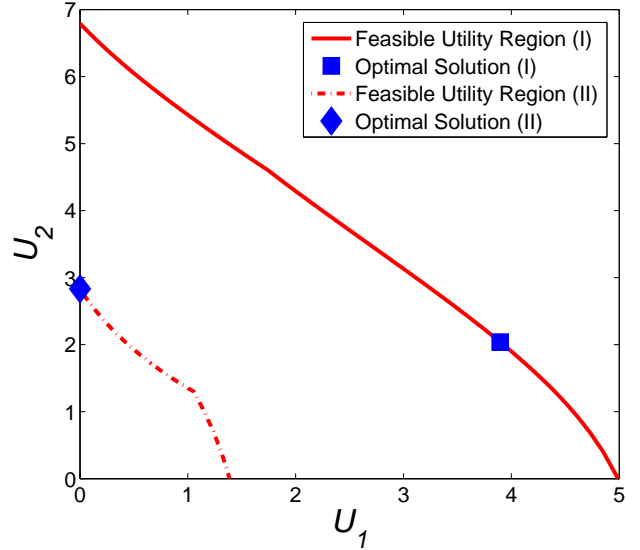}}\hspace{-0.0cm}\vspace{0cm}
\caption{The feasible utility region $\mathcal{F}$. Case (I): the channel gains are given by $h_{11}=0.73$, $h_{12}=0.04$, $h_{21}=0.03$, and $h_{22}=0.89$, and the maximum power are $P_1^{\max}=20$, $P_2^{\max}=100$; Case (II): the channel gains are given by $h_{11}=0.30$, $h_{12}=0.50$, $h_{21}=0.03$, and $h_{22}=0.80$, and the maximum power are $P_1^{\max}=1$, $P_2^{\max}=2$. In both cases, the noise power is 0.1 for each link, and the weights are $w_1=0.57$, $w_2=0.43$.}\label{fig:example}
\end{figure}

\begin{table}
\centering
\begin{tabular}{|c|c|c|c|c|c|}
\hline
\multirow{2}{*}{Approach} & \multicolumn{2}{c|}{Case I} & \multicolumn{2}{c|}{Case II}\\
\cline{2-5}
&Power  & Sum Rate & Power  & Sum Rate\\
\hline
GP & [20, 7.68] & 3.02 & [1, 0.61]& 0.98\\
\hline
ADP & [20, 6.46] & 3.10 & [1, 2]& 1.16\\
\hline
MAPEL & [20, 6.79] & 3.10 &[0, 2]& 1.22\\
\hline
SEER & [20, 6.90] & 3.10 &[0, 2]& 1.22\\
\hline
\end{tabular}
\caption{The performance of the existing approaches for Case I and II.}\label{tab:performance}
\end{table}

\emph{Example:} For the case with two links, Fig. \ref{fig:example} illustrates the nonconvex feasible utility region $\mathcal{F}$ for different system parameters. We compare the performance of the existing approaches \cite{chiang:2008,huang:2006,qian:2010,qian:2009} in Table \ref{tab:performance}.

\textbf{Remarks:} The solutions to (\ref{eq:sum}) given by \cite{chiang:2008,huang:2006,qian:2009} are either distributed but suboptimal or optimal but centralized. In particular, \cite{chiang:2008} solves (\ref{eq:sum}) by using geometric programming (GP) under the high-SINR assumption, which yields a suboptimal solution to (\ref{eq:sum}) when the assumption does not hold (e.g., this is the case in this example above). The ADP algorithm \cite{huang:2006} can guarantee only local optimality\footnote{The local optimal solution found by ADP happens to be globally optimal only in one of the cases that are illustrated in Table \ref{tab:performance}.} in a distributed manner. The MAPEL algorithm \cite{qian:2009} can achieve the globally optimal solutions but it is centralized with high computational complexity. Compared with these algorithms, the SEER algorithm \cite{qian:2010} can  guarantee global optimality in a distributed manner but message passing needed in each iteration can be demanding, i.e., each link needs the knowledge of the channel gains, the receiver SINR and the signal power of all the other links. It is worth noting that the performance of SEER hinges on the control parameter that can be challenging to choose on the fly.

\subsection{From Network Utility Maximization to Minimum Weighted Utility Maximization}

In order to devise low-complexity distributed algorithms that can guarantee global optimality, we first study the basic properties for the solutions to (\ref{eq:sum}), and then convert (\ref{eq:sum}) into a more structured max-min problem.

\begin{lm}
The optimal solution to (\ref{eq:sum}) is on the boundary of the feasible utility region $\mathcal{F}$.
\label{lm:boundary}
\end{lm}
\begin{proof}
Let $\mathbf{U}^*$ denote a globally optimal solution to (\ref{eq:sum}) over $\mathcal{F}$, and $\mathbf{\gamma}^*$ denote the corresponding SINR that supports $\mathbf{U}^*$. Since $U_l(\cdot)$ is continuous and strictly increasing,  proving that $\mathbf{U}^*$ is on the boundary of $\mathcal{F}$ is equivalent to showing that $\mathbf{\gamma}^*$ is on the boundary of the feasible SINR region. Suppose that $\mathbf{\gamma}^*$ is not on the boundary of the feasible SINR region, which indicates that there exists some point $\mathbf{\hat \gamma}$ such that $\hat \gamma_l \ge \gamma_l^*$ for all $l\in\mathcal{L}$ and $\hat \gamma_l > \gamma_l^*$ for some $l$. Since $U_l(\cdot)$ for any $l\in{\mathcal{L}}$ is strictly increasing in $\gamma_l$, we have $U_l(\hat \gamma_l) \ge U_l(\gamma_l^*)$ for all $l\in\mathcal{L}$ and $U_l(\hat \gamma_l) > U_l(\gamma_l^*)$ for some $l$, which contradicts the fact $\mathbf{\gamma}^*$ is a globally optimal solution. Hence, Lemma \ref{lm:boundary} follows.
\end{proof}

Based on Lemma \ref{lm:boundary}, if we can characterize the boundary of $\mathcal{F}$, then it is possible to solve (\ref{eq:sum}) efficiently. Thus motivated, we first establish, by introducing a ``contribution weight'' for each user, the equivalence between (\ref{eq:sum}) and the minimum weighted utility maximization problem.
\begin{lm}
Problem (\ref{eq:sum}) is equivalent to the following minimum weighted utility maximization:
\begin{equation}%
\begin{array}
[c]{lll}%
&{\text{maximize} }&\min_{l\in\mathcal{L}}\frac{U_l(\gamma_l(\mathbf{p}))}{x_l}
\\
&\text{subject to}& 0\leq p_l\leq P_l^{\max}, \forall l\in\mathcal{L}
\\
&& 0\leq x_l\leq 1, \forall l\in\mathcal{L}
\\
&& \sum_{l\in\mathcal{L}} x_l =1
\\
& \text{variables}& \{\mathbf{p},\mathbf{x}\}.
\end{array}
\label{eq:maxmin}
\end{equation}
\label{lm:maxmin}
\end{lm}
\begin{proof}
Let $t=\sum_{l\in\mathcal{L}}U_l(\gamma_l(\mathbf{p}))$ denote the total utility. Since $U_l(.)$ is nonnegative, we define $x_l\in[0,1]$ as a ratio for the contribution of user $l$'s utility to $t$. Therefore, $U_l(\gamma_l(\mathbf{p}))=tx_l$ and $\sum_{l\in\mathcal{L}} x_l =1$. Then (\ref{eq:sum}) can be rewritten as
\begin{equation}%
\begin{array}
[c]{lll}%
&{\text{maximize} }&t
\\
&\text{subject to}& t=\frac{U_l(\gamma_l(\mathbf{p}))}{x_l}, \forall l\in\mathcal{L}
\\
&&0\leq p_l\leq P_l^{\max}, \forall l\in\mathcal{L}
\\
&& 0\leq x_l\leq 1, \forall l\in\mathcal{L}
\\
&&0\le t, \sum_{l\in\mathcal{L}} x_l =1
\\
& \text{variables}& \{\mathbf{p},\mathbf{x},t\}.
\end{array}
\label{eq:fix_x}
\end{equation}
Then, in the context of maximizing $t$, it suffices to relax $t=\frac{U_l(\gamma_l(\mathbf{p}))}{x_l}$ in (\ref{eq:fix_x}) as $t\le\frac{U_l(\gamma_l(\mathbf{p}))}{x_l}, \;\forall l\in\mathcal{L}$, which is equivalent to $t\le\min_{l\in\mathcal{L}}\frac{U_l(\gamma_l(\mathbf{p}))}{x_l}$. Therefore, (\ref{eq:fix_x}) can be treated as the hypograph form of (\ref{eq:maxmin}), i.e., (\ref{eq:fix_x}) and (\ref{eq:maxmin}) are equivalent \cite{boyd:2004}, thereby concluding the proof.
\end{proof}

For given $\mathbf{x}$, (\ref{eq:maxmin}) is quasi-convex\footnote{By definition, a function $f : \mathbb{R}^n\rightarrow\mathbb{R}$ is quasi-convex, if its domain ${\textbf{dom}}f$ and all its sublevel sets $\mathcal{S}_c=\{x\in \text{\textbf{dom}}f|f(x)\le c\}$, for $c\in \mathbb{R}$, are convex \cite{boyd:2004}.}. By introducing an auxiliary variable $t$, we obtain the following equivalent formulation:
\begin{equation}%
\begin{array}
[c]{lll}%
&{\text{maximize} }&t
\\
&\text{subject to}& U_l^{-1}(tx_l)(n_l+\sum_{k\ne l}{h_{kl}p_k})\le h_{ll}p_l
\\
&& 0\leq p_l\leq P_l^{\max}, \forall l\in\mathcal{L}, \;0\le t
\\
& \text{variables}& \{\mathbf{p},t\},
\end{array}
\label{eq:bs}
\end{equation}
which can be solved in polynomial time through binary search on $t$ \cite{boyd:2004}. By transforming (\ref{eq:sum}) to this more structured max-min problem (\ref{eq:maxmin}), we are able to find each boundary point efficiently. Then, the problem is reduced to finding a globally optimal $\mathbf{x}^*$, given which we can obtain a globally optimal solution, i.e., the tangent point of the hyperplane and $\mathcal{F}$, as illustrated in Fig. \ref{fig:fr}. Intuitively speaking, $\mathbf{x}$ represents a search direction. Once we find the best search direction $\mathbf{x}^*$, $\mathbf{p}^*$ can be obtained efficiently by searching along the direction of $\mathbf{x}^*$.

\begin{figure}[t]
\begin{center}
\vspace{0.5cm}\hspace{-0cm} {\includegraphics[scale=0.5]{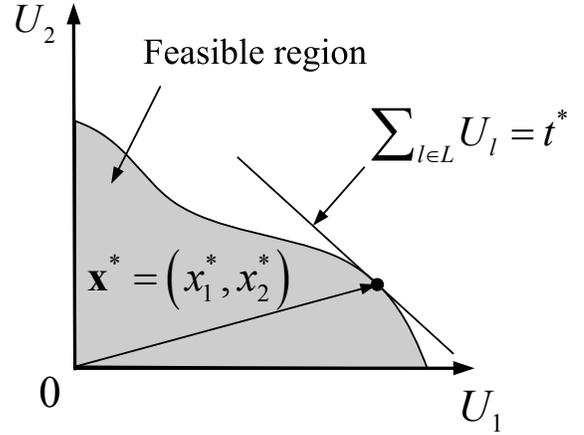}}\hspace{-0.0cm}
\vspace{5.5cm} \caption{An illustration of the max-min problem for the case with two links.}\vspace{-0.0cm}
\label{fig:fr}
\end{center}
\end{figure}

\subsection{Algorithms for Global Optimization}
In this section, we study algorithms achieving global optimality for (\ref{eq:maxmin}). First, we propose a centralized algorithm for (\ref{eq:maxmin}),
which will serve as a benchmark for performance comparison. Then, by using EDT and SA, we propose a distributed algorithm, DSPC, for the problem (\ref{eq:maxmin}). Building on this, we propose an enhanced algorithm EDSPC to improve the convergence rate of DSPC.
\\
\\
\noindent
{\emph{1) A Centralized Algorithm}

Based on Lemma \ref{lm:boundary} and Lemma \ref{lm:maxmin}, we develop a centralized algorithm (Algorithm \ref{alg:bsa}) to solve the max-min optimization problem (\ref{eq:maxmin}) under consideration. Roughly speaking, by dividing the simplex $\mathcal{S}=\{\mathbf{x}|\sum_{l\in\mathcal{L}} x_l =1, 0\leq x_l\leq 1, \forall l\in\mathcal{L}\}$ into many small simplices, the algorithm can find the optimal point on the boundary of $\mathcal{F}$.

\begin{prop}
Algorithm \ref{alg:bsa} converges monotonically to a globally optimal solution to (\ref{eq:maxmin}) as the approximation factor $\epsilon$ approaches zero.
\end{prop}
\begin{proof}
For given $\epsilon$, Algorithm \ref{alg:bsa} divides the simplex $\mathcal{S}=\{\mathbf{x}|\sum_{l\in\mathcal{L}} x_l =1, 0\leq x_l\leq 1, \forall l\in\mathcal{L}\}$ into $\lceil1/\epsilon\rceil$ simplices\footnote{$\lceil1/\epsilon\rceil$ denotes the smallest integer greater than $1/\epsilon$.}. Then Algorithm \ref{alg:bsa} computes the power allocation $\mathbf{p^*}$ by solving (\ref{eq:bs}) at $\mathbf{x}$ given by the center point of the simplex. Since the optimal search direction $\mathbf{x}^*$ is in $\mathcal{S}$, when the approximation factor $\epsilon$ approaches zero, Algorithm \ref{alg:bsa} exhaustively searches every point in the simplex $\mathcal{S}$. Therefore, Algorithm \ref{alg:bsa} can converge monotonically to a globally optimal solution to (\ref{eq:maxmin}).
\end{proof}

{\bf Remarks:} In Algorithm \ref{alg:bsa},   by controlling $\epsilon$, one can obtain a solution arbitrarily close to a globally optimal one.
  Accordingly, Algorithm \ref{alg:bsa} can guarantee global optimality. However, the complexity of this algorithm can be high and this is possible only with centralized implementation. Algorithm 1 will be used only as a benchmark for performance evaluation of distributed algorithms.

\begin{algorithm}
\caption{}
\label{alg:bsa}
\begin{algorithmic}
\STATE \textbf{Initialization}: Choose the approximation factor $\epsilon>0$, and construct the initial simplex $S$ with the vertex set $V=\{v_1,...,v_L\}$, where $v_l=e_l$ and $e_l$ is the $l$th unit coordinate vector. Let $v_c=\frac{1}{L}\sum_{l\in\mathcal{L}}v_l$ be the center of $S$. Compute $\mathbf{p^*}$  by solving (\ref{eq:bs}) at the point $\mathbf{x} = v_c$. 
\STATE \textbf{Repeat}
\begin{enumerate}
\STATE Each simplex is divided into $L$ subsimplices $S_1,...,S_L$. Let $S$ be a simplex with the vertex set $V=\{v_1,...,v_L\}$, and let $v\in S\backslash V$. Choose $v=\frac{1}{L}\sum_{l\in\mathcal{L}}v_l$. Then, each simplex $S_l$ is defined as having vertex set $V\backslash v_l\bigcup v$.
\STATE For each new simplex, compute $\mathbf{p^*}$ by solving (\ref{eq:bs}) at $\mathbf{x}$ given by the center point of the simplex.
\STATE Find the current best solution to (\ref{eq:maxmin}).
\end{enumerate}
\STATE \textbf{Until} The number of simplices is greater than $1/{\epsilon}$.
\end{algorithmic}
\end{algorithm}

\noindent
{\emph{2) DSPC Algorithm}

Next, we devise a distributed stochastic power control (DSPC) algorithm based on EDT \cite{chen:2008} and SA \cite{Kirkpatrick:1983}. To this end, we first introduce auxiliary variables and use EDT to transform (\ref{eq:maxmin}) with the auxiliary variables into an unconstrained problem. Then, we solve the unconstrained problem by using the SA mechanism. Specifically, define $t_l=\frac{U_l(\gamma_l(\mathbf{p}))}{x_l}$ and rewrite (\ref{eq:maxmin}) as
\begin{equation}%
\begin{array}
[c]{lll}%
&{\text{minimize} }& -\min_{l\in\mathcal{L}}t_l
\\
&\text{subject to}& t_lx_l\le U_l(\gamma_l(\mathbf{p})), \forall l\in\mathcal{L}
\\
&& \sum_{l\in\mathcal{L}} x_l =1
\\
&& 0\leq p_l\leq P_l^{\max}, \forall l\in\mathcal{L}
\\
&& 0\le t_l, 0\leq x_l\leq 1, \forall l\in\mathcal{L}
\\
& \text{variables}& \{\mathbf{p},\mathbf{x},\mathbf{t}\}.
\end{array}
\label{eq:maxmin2}
\end{equation}

Next, we use EDT to write Lagrangian for (\ref{eq:maxmin2}) as
\begin{equation}%
\begin{array}
[c]{lll}%
L(\mathbf{p},\mathbf{x},\mathbf{t},\alpha,\boldsymbol{\beta})&=&-\min_{l\in\mathcal{L}}t_l+\alpha\left|\sum_{l\in\mathcal{L}} x_l-1\right|\\
&&+\sum_{l\in\mathcal{L}}\beta_l(t_lx_l- U_l(\gamma_l(\mathbf{p})))^{+},
\end{array}
\label{eq:penalty_maxmin}
\end{equation}
where $(y)^{+}=\max(0,y)$, and $\alpha\in\mathbb{R}$ and $\boldsymbol{\beta}\in\mathbb{R}^L$ are the penalty multipliers for penalizing the constraint violations. Based on EDT \cite{chen:2008}, there exist finite $\alpha^*\ge 0$ and $\beta_l^*\ge 0$ for all $l\in\mathcal{L}$ such that, for any $\alpha>\alpha^*$ and $\beta_l>\beta_l^*$, $\forall~l\in\mathcal{L}$, the solution to (\ref{eq:penalty_maxmin}) is the same as (\ref{eq:maxmin2}). Note that (\ref{eq:penalty_maxmin}) does not include the power constraints, due to the fact that there is no coupling among the user powers. Therefore, the minimization of (\ref{eq:penalty_maxmin}) with respect to the primal variables ($\mathbf{p}$, $\mathbf{x}$, and $\mathbf{t}$) can be carried out individually by each user in a distributed fashion.

The next key step is to perform a stochastic local search by each user based on SA. Let $t_l$, $x_l$ and $p_l$ denote the primal values of the $l$th user, and $t_l'$ and $x_l'$ denote the new values randomly chosen by the $l$th user. Accordingly, $t_l'x_l'$ can be treated as a new target utility for the $l$th user. To achieve this target utility, the $l$th user updates $p_l'$ by
\begin{equation}%
\begin{array}
[c]{lll}%
p_l'=\min\left(\frac{U_l^{-1}(t_l'x_l')}{\gamma_l}p_l,P_l^{\max}\right),
\end{array}
\label{eq:pc}
\end{equation}
where $\gamma_l$ is the current SINR measured at the $l$th user's receiver. Note that (\ref{eq:pc}) does not need any information of channel gains except the SINR feedback, i.e., $\gamma_l$. Since (\ref{eq:pc}) corresponds to the distributed power control algorithm of standard form as described in \cite{yates:1995}\footnote{A power control algorithm is of standard form, if the interference function (the effective interference each link must overcome) is positive, monotonic and scalable in power allocation \cite{yates:1995}.}, it converges geometrically fast to the target utility. Thus, we assume that each user $l$ updates $p_l$ in a faster time-scale than $t_l$ and $x_l$ such that $p_l$ always converges before the next update of $t_l$ and $x_l$.
Let $\Delta$ denote the difference between $L(p_l,x_l,t_l|\mathbf{p}_{-l},\mathbf{x}_{-l},\mathbf{t}_{-l},\alpha,\boldsymbol{\beta})$ and $L(p_l',x_l',t_l'|\mathbf{p}_{-l}',\mathbf{x}_{-l},\mathbf{t}_{-l},\alpha,\boldsymbol{\beta})$, where $\mathbf{y}_{-l}$ is the vector $\mathbf{y}$ without the $l$th user's variable. If $\Delta\ge 0$, i.e., $t_l'$, $x_l'$ and $p_l'$ reduce Lagrangian (\ref{eq:penalty_maxmin}), then they are accepted with probability 1; otherwise, they are accepted with probability $\exp\left(\frac{\Delta}{T}\right)$, where $T$ is a control parameter (it is also called temperature). Note that, as $T$ decreases, the acceptance of uphill move becomes less and less probable, and therefore a fine-grained search takes place. It has been shown that, as $T$ tends to 0 according to a \emph{logarithmic cooling schedule}, SA converges to a globally optimal point \cite{geman:1984,hajek:1988}. To compute $\Delta$ locally by each user $l$, we assume that user $l$ broadcasts the terms $t_l$, $x_l$ and $\beta_l(t_lx_l- U_l(\gamma_l(\mathbf{p})))^{+}$, whenever any of these terms changes.

Besides updating the primal variables, each user $l$ also needs to update $\alpha$ and $\beta_l$ to satisfy $\alpha>\alpha^*$ and $\beta_l>\beta^*_l$. Here, we apply the method given by \cite{chen:2008} to update $\alpha$ and $\beta_l$. In particular, if any constraint is violated, $\alpha$ and $\beta_l$ are updated as follows:
\begin{equation}%
\begin{array}
[c]{lll}%
\alpha&\leftarrow& \alpha+\sigma\left|\sum_{l\in\mathcal{L}} x_l-1\right|,
\\
\beta_l&\leftarrow& \beta_l+\varrho_l(t_lx_l- U_l(\gamma_l))^{+}, \; \forall l\in \mathcal{L},
\end{array}
\label{eq:penalty_variables}
\end{equation}
where $\sigma$ and $\varrho_l$ are used to control the rate of updating $\alpha$ and $\beta_l$. Thus, after initialization, $\alpha$ and $\beta_l$ increase in proportion to the violation of the corresponding constraint, which may lead to excessively large penalty values. Since it is beneficial to periodically scale down the penalty values to ease the unconstrained optimization, $\alpha$ and $\beta_l$ are scaled down by multiplying with a random value (it is chosen empirically between 0.7 to 0.95){\footnote{See \cite{chen:2008} for the detailed description on the choice of these parameters.\label{fn:scaledown}}} if the \emph{penalty decrease condition} is satisfied, i.e., the maximum violation of constraints is not decreased after running Step 1 in Algorithm \ref{alg:dspc} several times consecutively, e.g., five times$^\text{\ref{fn:scaledown}}$. A detailed description of DSPC algorithm is given in Algorithm \ref{alg:dspc}.

\begin{prop}
The distributed stochastic power control algorithm (Algorithm \ref{alg:dspc}) converges monotonically to a globally optimal solution to (\ref{eq:maxmin}), as temperature $T$ in SA decreases to zero.
\label{prop:dspc}
\end{prop}
\begin{proof}
For a given pair of $\alpha$ and $\boldsymbol{\beta}$, Algorithm \ref{alg:dspc} converges to a globally optimal solution to (\ref{eq:penalty_maxmin}) by using the logarithmic cooling schedule \cite{geman:1984,hajek:1988}. If the solution satisfies the constraints of (\ref{eq:maxmin2}), it is also a globally optimal solution to (\ref{eq:maxmin2}) based on EDT, i.e., current $\alpha$ and $\boldsymbol{\beta}$ satisfy $\alpha>\alpha^*$ and $\beta_l>\beta^*_l$ for all $l\in{\mathcal{L}}$ \cite{chen:2008}. By iteratively updating $\alpha$ and $\boldsymbol{\beta}$, Algorithm \ref{alg:dspc} will converge to a globally optimal solution to (\ref{eq:maxmin}), when $\alpha$ and $\boldsymbol{\beta}$ satisfy $\alpha>\alpha^*$ and $\beta_l>\beta^*_l$ for all $l\in{\mathcal{L}}$.
\end{proof}

{\bf Remarks:} The DSPC algorithm can guarantee global optimality in a distributed manner without the need of channel information. In particular, it can adapt to channel variations by utilizing the SINR feedback. However, the convergence rate of DSPC is slow due to the use of logarithmic cooling schedule.

\begin{algorithm}
\caption{Distributed Stochastic Power Control (DSPC)}
\label{alg:dspc}
\begin{algorithmic}
\STATE \textbf{Initialization}: Choose $\epsilon>0$. Let $\alpha=0$, $\beta_l=0$, $\forall l\in\mathcal{L}$, and randomly choose $\mathbf{p}$, $\mathbf{x}$ and $\mathbf{t}$.
\STATE \textbf{Step 1: update primal variables}
\STATE Set $T=T_0$, and select a sequence of time epochs $\{\tau_1,\tau_2,... \}$ in continuous time.
\STATE \quad \textbf{Repeat for each user} $l$
\begin{enumerate}
\STATE Randomly pick $t_l'$ and $x_l'$ in the feasible region, and update $p_l'$ according to (\ref{eq:pc}).
\STATE Keep sensing the change of $\beta_l(t_lx_l- U_l(\gamma_l(\mathbf{p})))^{+}$ broadcast by other users.
\STATE Compute $\Delta$, and accept $t_l'$, $x_l'$, and $p_l'$ with probability 1, if $\Delta\ge 0$, or with probability $\exp(\frac{\Delta}{T})$, otherwise.
\STATE Broadcast $t_l'$ and $x_l'$, if $t_l'$ and $x_l'$ are updated.
\STATE For each time epoch $\tau_i$, update $T=T_0/\log(i+1)$.
\end{enumerate}
\STATE \quad \textbf{Until} $T<\epsilon$.
\STATE \textbf{Step 2: update penalty variables}
\STATE \quad \textbf{For each user} $l$,
\begin{enumerate}
\STATE Update $\alpha$ and $\beta_l$ according to (\ref{eq:penalty_variables}), and scale down $\alpha$ and $\beta_l$, if the penalty decrease condition is satisfied.
\STATE Goto Step 1 until no constraint is violated.
\end{enumerate}
\end{algorithmic}
\end{algorithm}

\noindent
\\
\emph{3) Enhanced DSPC Algorithm}

It can be seen from Algorithm \ref{alg:dspc} that it is critical to find the optimal penalty variables $\alpha$ and $\boldsymbol{\beta}$ for computing (\ref{eq:penalty_maxmin}). Moreover, a logarithmic cooling schedule is used to ensure convergence to a global optimum. To improve the convergence rate, we propose next an enhanced algorithm for DSPC (EDSPC) by empirically choosing the initial penalty values $\alpha_0$ and $\boldsymbol{\beta}_0$ and employing a \emph{geometric cooling schedule} \cite{Kirkpatrick:1983}, which reduces the temperature $T$ in SA by $T=\xi T$, $0<\xi<1$, at each time epoch. Compared with the logarithmic cooling schedule, $T$ converges to 0 much faster under the geometric cooling schedule, which in turn improves the convergence rate of DSPC. The resulting solution is given in Algorithm \ref{alg:hdspc}.

We note that although EDSPC converges much faster than DSPC, it may yield only near-optimal solutions. Based on EDT, we  choose $\alpha_0>\alpha^*$ and $\beta_{0l}>\beta_l^*,~\forall~l\in\mathcal{L}$, to satisfy the optimality conditions for penalty variables. Obviously, by choosing large $\alpha_0$ and $\beta_{0l}$, these conditions can be always satisfied. Nevertheless, very large penalties introduce heavy costs for constraint violations such that EDSPC may end up with a feasible but suboptimal solution. Therefore, the selection of initial penalty values plays a critical role in  the performance of EDSPC and deserves more attention in future work.


\begin{algorithm}
\caption{Enhanced Distributed Stochastic Power Control (EDSPC)}
\label{alg:hdspc}
\begin{algorithmic}
\STATE \textbf{Initialization}: Choose $\epsilon>0$. Let $\alpha=\alpha_0$, $\beta_l=\beta_{0l}$, $\forall l\in\mathcal{L}$, and randomly choose $\mathbf{p}$, $\mathbf{x}$ and $\mathbf{t}$.
\STATE Set $T=T_0$, and select a sequence of time epochs $\{\tau_1,\tau_2,... \}$ in continuous time.
\STATE \textbf{Repeat for each user} $l$
\begin{enumerate}
\STATE Randomly pick $t_l'$ and $x_l'$ in the feasible region, and update $p_l'$ according to (\ref{eq:pc}).
\STATE Keep sensing the change of $\beta_l(t_lx_l- U_l(\gamma_l(\mathbf{p})))^{+}$ broadcast by other users.
\STATE Compute $\Delta$, and accept $t_l'$, $x_l'$, and $p_l'$ with probability 1, if $\Delta\ge 0$, or with probability $\exp(\frac{\Delta}{T})$, otherwise.
\STATE Broadcast $t_l'$ and $x_l'$, if $t_l'$ and $x_l'$ are updated.
\STATE For each time epoch $\tau_i$, update $T=\xi T$.
\end{enumerate}
\STATE \textbf{Until} $T<\epsilon$.
\end{algorithmic}
\end{algorithm}

\subsection{Numerical Examples}\label{sec:unicastnumerical}

In this section, we evaluate the utility and convergence performance of Algorithms \ref{alg:dspc} and \ref{alg:hdspc} (DSPC\footnote{The geometric cooling schedule is employed to accelerate the convergence rate of DSPC in the simulation. DSPC updates penalty values until they satisfy the threshold-based optimality condition.} and EDSPC). We consider a wireless network with six links randomly distributed on a 10m-by-10m square area. The channel gains $h_{lk}$ are equal to $d_{lk}^{-4}$, where $d_{lk}$ represents the distance between the transmitter of user $l$ and the receiver of user $k$. We assume  $U_l(\gamma_l(\mathbf{p}))=\log(1+\gamma_l(\mathbf{p}))$, $P_{l}^{\max}=1$ and $n_l=10^{-4}$ for all $l\in\mathcal{L}$, and consider one randomly generated realization of channel gains given by

\vspace{-0.25cm}
\begin{small}
\[
\mathbf{H }= \left[ {\begin{array}{*{20}c}
   {{\rm{0}}{\rm{.3318}}} & {{\rm{0}}{\rm{.0049}}} & {{\rm{0}}{\rm{.0141}}} & {{\rm{0}}{\rm{.0021}}} & {{\rm{0}}{\rm{.0016}}} & {{\rm{0}}{\rm{.0007}}}  \\
   {{\rm{0}}{\rm{.0031}}} & {{\rm{0}}{\rm{.9554}}} & {{\rm{0}}{\rm{.0063}}} & {{\rm{0}}{\rm{.0140}}} & {{\rm{0}}{\rm{.0012}}} & {{\rm{0}}{\rm{.0025}}}  \\
   {{\rm{0}}{\rm{.0155}}} & {{\rm{0}}{\rm{.0042}}} & {{\rm{0}}{\rm{.6166}}} & {{\rm{0}}{\rm{.0046}}} & {{\rm{0}}{\rm{.0108}}} & {{\rm{0}}{\rm{.0018}}}  \\
   {{\rm{0}}{\rm{.0017}}} & {{\rm{0}}{\rm{.2188}}} & {{\rm{0}}{\rm{.0340}}} & {{\rm{0}}{\rm{.6754}}} & {{\rm{0}}{\rm{.0062}}} & {{\rm{0}}{\rm{.0215}}}  \\
   {{\rm{0}}{\rm{.0020}}} & {{\rm{0}}{\rm{.0017}}} & {{\rm{0}}{\rm{.2216}}} & {{\rm{0}}{\rm{.0042}}} & {{\rm{0}}{\rm{.2955}}} & {{\rm{0}}{\rm{.0028}}}  \\
   {{\rm{0}}{\rm{.0007}}} & {{\rm{0}}{\rm{.0079}}} & {{\rm{0}}{\rm{.0254}}} & {{\rm{0}}{\rm{.2553}}} & {{\rm{0}}{\rm{.0404}}} & {{\rm{0}}{\rm{.3025}}}  \\
\end{array}} \right].
\]
\end{small}

Fig. \ref{fig:convergence} shows how the total utility in the EDSPC algorithm converges over time, where we choose all the initial penalty values equal to 10. Also, we choose $\xi=0.9$, $\rho=1$ and $\varrho=1$, and use Algorithm \ref{alg:bsa} as a benchmark to evaluate the optimal performance. As shown in Fig. \ref{fig:convergence}, the EDSPC algorithm  approaches the optimal utility, when the initial penalty values are carefully chosen. Moreover, the convergence rate of the EDSPC algorithm is much faster than DSPC, since DSPC continues updating the penalty values after the optimal solution is found for the current penalty values. Fig. \ref{fig:comparison} illustrates the average performance (with confidence interval) of DSPC, EDSPC, and SEER under 100 random initializations, with the same system parameters as in Fig. \ref{fig:convergence}. As shown in Fig. \ref{fig:comparison}, both DSPC and EDSPC are robust against the initial value variations.

\begin{figure}[t]
\begin{center}
\vspace{-0.0cm}\hspace{-0.0cm} {\includegraphics[scale=0.45]{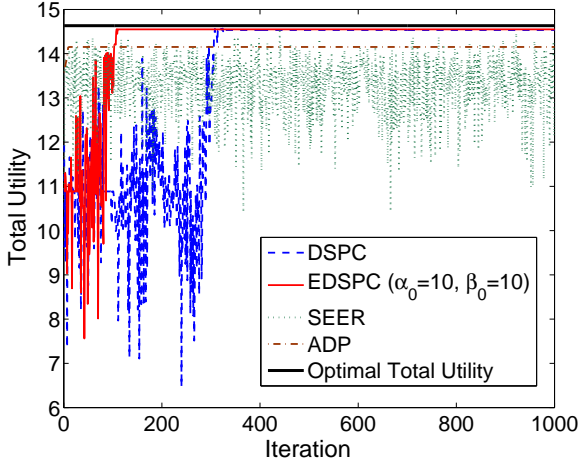}}\hspace{-0.0cm}
\vspace{0cm} \caption{Convergence performance of DSPC, EDSPC, SEER and ADP.}\vspace{-0.0cm}
\label{fig:convergence}
\end{center}
\end{figure}

\begin{figure}[t]
\begin{center}
\vspace{-0.0cm}\hspace{0cm} {\includegraphics[scale=0.45]{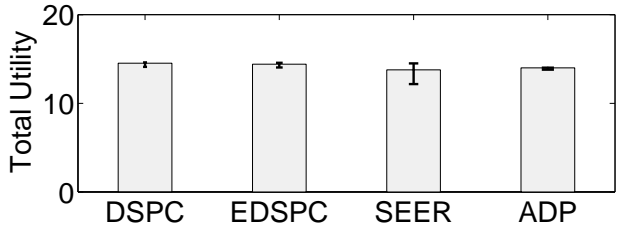}}\hspace{-0cm}
\vspace{0cm} \caption{Comparison of the average utility performance (with confidence interval) of DSPC, EDSPC, SEER and ADP.}\vspace{-0.0cm}
\label{fig:comparison}
\end{center}
\end{figure}

A comparison with the SEER and ADP is also depicted in Fig. \ref{fig:convergence} and \ref{fig:comparison}. As mentioned in Section \ref{sec:introduction}, ADP can only guarantee local optimality. Therefore, for nonconvex problems (e.g., in this example), ADP may converge to a suboptimal solution. As noted in \cite{qian:2010}, the performance of SEER heavily hinges on the control parameter that can be challenging to choose on the fly. In contrast, DSPC can approach the globally optimal solution regardless of the initial parameter selection, but the convergence rate may be slower. Further, EDSPC improves the convergence rate, but the initial penalty values would impact how close it can approach the optimal point. From the point of view of reducing the amount of message passing, in our algorithms each link does not need any knowledge of the channel gains (including its own channel gain), the receiver SINR of the other links and the signal power of the other links, which are all used in the SEER algorithm.


\section{Joint Scheduling and Power Control for Stability of Queueing Systems}\label{sec:stability}
In Section \ref{sec:pc_unicast}, we studied the distributed power allocation, by using DSPC and EDSPC, for utility maximization in the saturated case. In this section, we generalize the study by considering a queueing system with dynamic packet arrivals and departures. Specifically, we develop a joint scheduling and power allocation policy to stabilize packet queues by integrating our power control algorithms with the celebrated back-pressure algorithm\cite{ephremides:1992}.

\subsection{Stability Region and Throughput Optimal Power Allocation Policy}
Consider the same wireless network model with $L$ links as in Section \ref{sec:pc_unicast}. We assume that there are $S$ classes of users in the system, and that the traffic brought by users of class $s$ follows $\{A_{sl}(t)\}_{t=1}^\infty$, which are i.i.d. sequences of random variables for all $l=1,...,L$ and $s=1,...,S$, where $A_{sl}(t)$ denotes the amount of traffic generated by users of class $s$ that enters the link $l$ in slot $t$. Let $Q_{T(l)}^s(t)$ and $Q_{R(l)}^s(t)$ denote the current backlog in the queue of class $s$ in slot $t$ on the transmitter and receiver sides of link $l$, respectively. The queue length $Q_{T(l)}^s(t)$ evolves over time as
\begin{equation}%
\begin{array}
[c]{lll}%
Q_{T(l)}^s(t+1)&=&\max(Q_{T(l)}^s(t)-r_{l}^s(t),0)+A_{sl}(t)\\
&&+\sum_{\{m|T(l)=R(m),m\in\mathcal{L}\}}r_m^s(t),
\end{array}
\label{eq:queue}
\end{equation}
where $r_{l}^s(t)$ denotes the transmission rate of link $l$ for users of class $s$. The third term in (\ref{eq:queue}) denotes the traffic from the other links. Assuming that the second moments of the arrival process $\{A_{sl}(t)\}_{t=1}^\infty$ are finite, the queue length process $\{Q_{T(l)}^s(t)\}_{t=1}^\infty$ forms a Markov chain.

Let $E_{sl}=1$ be the indicator that the path of users of class $s$ uses link $l$, and $E_{sl}=0$, otherwise. As is standard \cite{lin:2008,ephremides:1992,neely:2005}, the stability region is defined as follows.
\begin{defn}
The stability region $\Lambda$ is the closure of the set of all $\{\psi_s\}_{s=1}^S$ for which there exists some feasible power allocation policy under which the system is stable, i.e., $\Lambda=\bigcup_{\mathbf{p}\in\mathcal{P}}{\Lambda(\mathbf{p})}$, where $\Lambda(\mathbf{p})=\{ \{\psi_s\}_{s=1}^S | \sum_{s=1}^S E_{sl}\psi_s < r_l(\mathbf{p}), \forall l\}$, and $\mathcal{P}$ denotes the set of feasible power allocation. Here $\psi_s$ denotes the first moment of $\{A_{sl}(t)\}_{t=1}^\infty$, i.e., the load brought by users of class $s$, and $r_l(\mathbf{p})$ denotes the rate of link $l$ under power allocation $\mathbf{p}$.
\end{defn}

For the sake of comparison, the throughput region\footnote{Note that the feasible utility region $\mathcal{F}$ defined in Section \ref{sec:pc_unicast} is the throughput region, when the utility function is the same as the rate function.} $\mathcal{F}$ of the corresponding saturated case is defined as the set of all feasible link rates, i.e., $\mathcal{F}=\{\mathbf{r}|r_l=r_l(\mathbf{p}), \mathbf{p}\in\mathcal{P}\}$. In general, the throughput region $\mathcal{F}$ may be different from the stability region $\Lambda$, except for some special cases (e.g., in slotted ALOHA systems the throughput region and the stability region are the same \cite{rao:1988} for two links and in a multiple-access channel the information theoretic capacity region is equivalent to its stability region \cite{parande:2008}).

The system is stable if the arrival rates of packet queues are less than the service rates such that the queue lengths do not grow to infinity. In order to stabilize packet queues, it is critical to find the optimal scheduling and power allocation policy that maximizes the weighted sum rate given by (\ref{eq:mtp}). By integrating our power control algorithms and the back-pressure algorithm, we propose the following joint scheduling and power allocation policy (presented in Algorithm \ref{alg:mtp}) to stabilize the queueing system.

\begin{prop}
The joint scheduling and power allocation policy (Algorithm \ref{alg:mtp}) can stabilize the system when the load $\{\psi_s\}_{s=1}^S$ is strictly interior to the stability region $\Lambda$, i.e., there exists some $\epsilon>0$ such that $\{\psi_s+\epsilon\}_{s=1}^S\in\Lambda$.
\end{prop}

The proof is similar to that in \cite{neely:2005,neely:2006}, and is omitted for brevity.

Note that Algorithm \ref{alg:mtp} can be viewed as a single-hop  dynamic back-pressure and resource allocation policy \cite{neely:2006},
crafted towards solving the weighted sum rate maximization problem (\ref{eq:mtp}). Specifically,
by using the DSPC algorithm, Algorithm \ref{alg:mtp} can be implemented distributively to find the globally optimal resource allocation.
We should caution that EDSPC can be applied to improve the convergence rate of Stage 2 in Algorithm \ref{alg:mtp} but it may render a suboptimal schedule (i.e., it can not stabilize all possible $\{\psi_s\}_{s=1}^S$ within $\Lambda$), due to the fact that EDSPC may not always find the global optimal power allocation.

To reduce the complexity, we can consider a policy that computes (\ref{eq:mtp}) every few slots, and it can be shown that this policy can also stabilize the system, when $\{\psi_s\}_{s=1}^S$ is strictly interior to the stability region $\Lambda$ \cite{sarkar:2008,zhang:2009}.

\begin{algorithm}
\caption{Joint Scheduling and Power Allocation Policy}
\label{alg:mtp}
\begin{algorithmic}
\STATE \textbf{Stage 1:} For each link $l$, select a link weight according to $w_l(t)=\max\limits_{s=1,...,S}D_{l}^s(t)$, where the difference of queue lengths of class $s$ is $D_{l}^s(t)=\max(Q_{T(l)}^s(t)-Q_{R(l)}^s(t),0)$, if the receiver of link $l$ is not the destination of class $s$'s traffic, and  $D_{l}^s(t)=Q_{T(l)}^s$, otherwise.

\STATE \textbf{Stage 2:} Compute the optimal power allocation $\mathbf{p}^*$ in each slot $t$ by solving the following problem with DSPC algorithm
\begin{equation}%
\begin{array}
[c]{lll}%
\mathbf{p}^*=\text{arg}\max\limits_{\mathbf{p}} \sum\limits_{l=1}^L w_l(t)r_l(\mathbf{p}).
\end{array}
\label{eq:mtp}
\end{equation}
Thus, the transmission rate of link $l$ in slot $t$ is given by $r_l(\mathbf{p}^*)=\log(1+\gamma_l(\mathbf{p}^*))$.

\STATE \textbf{Stage 3:} Let $s_l^*=\text{arg}\max\limits_{s=1,...,S}D_{sl}(t)$ denote the class scheduled in slot $t$; if multiple classes satisfy this condition, then $s_l^*$ is randomly chosen as one of these classes. Then, schedule these classes according to the solution given by Stage 2.
\end{algorithmic}
\end{algorithm}

\subsection{Numerical Examples}
In this section, we present numerical results to illustrate the use of Algorithm \ref{alg:mtp} for stabilizing a queueing system. We consider a one-hop network (i.e., $\mathbf{E}=\{E_{sl}\}$ is the identity matrix) with two users (classes), where the channel gains are given by $h_{11}=0.3$, $h_{12}=0.5$, $h_{21}=0.03$, and $h_{22}=0.8$, and the noise power is 0.1 for each link. The maximum transmission power is set to 1 and 2 for links 1 and 2, respectively. Besides, we assume that the users of class $s$ arrive at the network according to a Poisson process with rate $\lambda_s$, and that the size of file brought by each user follows an exponential distribution with mean $\nu_s$. The load brought by users of class $s$ is then $\psi_s=\lambda_s\nu_s$. For this example, we also study the stability region $\Lambda$ and compare it with the throughput region $\mathcal{F}$ of the corresponding saturated case as illustrated in Fig. \ref{fig:stability region}. The stability region follows from the union of link rates that are conditioned on whether the other link is backlogged or not \cite{rao:1988,parande:2008}. First, we derive the stability region for the given power allocation. Then, we vary power allocation in the feasible region, and by taking the envelope of these regions, we obtain the overall stability region shown in Fig. \ref{fig:stability region}. However, different from the previous cases, where the throughput region is the same as the stability region, e.g., in a slotted ALOHA system with two links \cite{rao:1988} and in a multiple-access channel \cite{parande:2008}, in our case under the SINR model, the throughput region $\mathcal{F}$ is strictly smaller than the stability region (due to the underlying nonconvex optimization problem), as observed from Fig. \ref{fig:stability region}, which is the convex hull of $\mathcal{F}$, i.e., $Co(\mathcal{F})$, achievable by timesharing across different transmission modes\footnote{The transmission mode is defined as the transmission rate pair within the throughput region $\mathcal{F}$.}.

\begin{figure}[t]
\begin{center}
\vspace{-0.0cm}\hspace{0cm} {\includegraphics[scale=0.45]{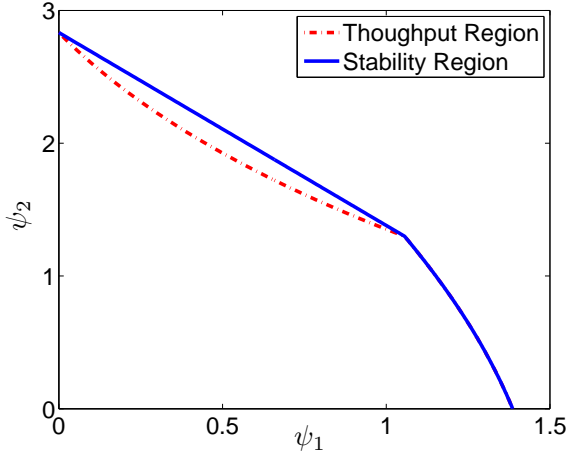}}\hspace{-0cm}
\vspace{0cm} \caption{Comparison of the stability region and the throughput region.}\vspace{-0.0cm}
\label{fig:stability region}
\end{center}
\end{figure}

\begin{figure}[t]
\begin{center}
\vspace{-0.0cm}\hspace{0cm} {\includegraphics[scale=0.45]{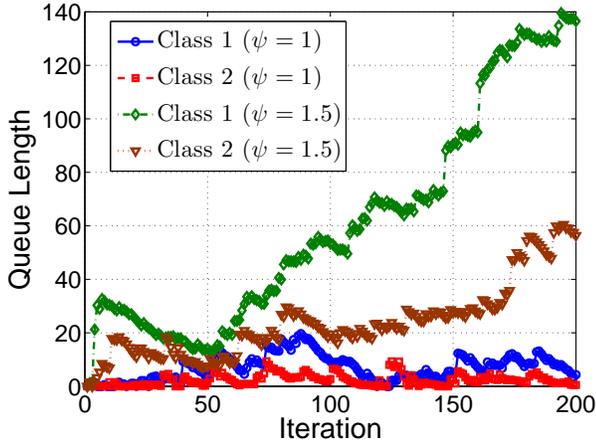}}\hspace{-0cm}
\vspace{0cm} \caption{Comparison of sample paths of a user's queue length for different traffic loads.}\vspace{-0.0cm}
\label{fig:queue}
\end{center}
\end{figure}

Then, we vary the arrival rate $\lambda$ and the average file size $\nu$ to change the traffic intensity $\psi=\lambda\nu$. Assuming that the arrival rate and the average file size of each user are the same, we compare the sample paths of each user's queue length for $\psi=1$ ($\lambda=1$, $\nu=1$) with $\psi=1.5$ ($\lambda=1.5$, $\nu=1$) in Fig. \ref{fig:queue}. When $\psi=1$, which falls in the stability region shown in Fig. \ref{fig:stability region}, the system is stabilized by using Algorithm \ref{alg:mtp}, while, when $\psi=1.5$, which is outside the stability region, the system becomes unstable. Fig. \ref{fig:delay} illustrates the average delay of the system as a function of the arrival rates. The delay is finite for small loads and grows unbounded when the loads are outside the stability region.

\begin{figure}[t]
\begin{center}
\vspace{-0.0cm}\hspace{0cm} {\includegraphics[scale=0.45]{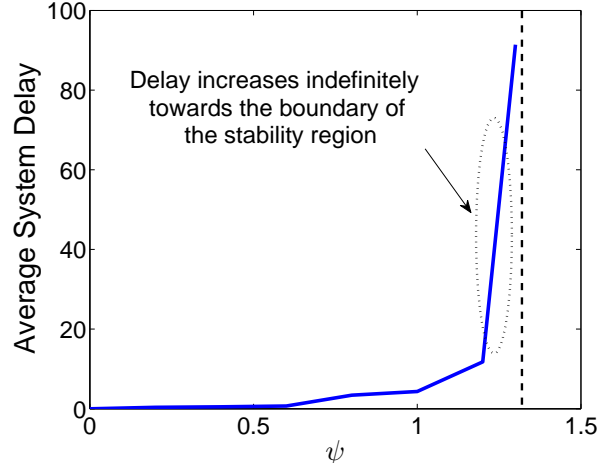}}\hspace{-0cm}
\vspace{0cm} \caption{Average delay of the system vs. system loads.}\vspace{-0.0cm}
\label{fig:delay}
\end{center}
\end{figure}


\section{Power control for Multicast Communications}\label{sec:multicast}
Due to wireless multicast advantage \cite{wieselthier:2000}, multicasting enables efficient data delivery to multiple recipients with a single transmission. In this section, we extend the distributed stochastic power control algorithms in Section \ref{sec:pc_unicast} to support multicast communications.

\subsection{System Model}
Beyond the model described in Section \ref{sec:pc_unicast}, we consider that each user $l$ has one transmitter and a set $\mathcal{M}_l$ of receivers. The corresponding transmission rate, $r_l$, is determined by the bottleneck link among these transmitter-receiver pairs, i.e., $r_l=\min_{m\in\mathcal{M}_l}r_{lm}$, where $r_{lm}$ denotes the link rate between the transmitter of user $l$ and its receiver $m$, and it is calculated based on the Shannon rate $\log(1+\gamma_{lm}(\mathbf{p}))$ for Gaussian, flat fading channels. Here, we do not consider the general broadcast capacity region but rather focus on maximizing the bottleneck link rates.

\subsection{Network Utility Maximization}
We seek to find the optimal power allocation $\mathbf{p}^*$ that maximizes the overall system utility subject to the power constraints in multicast communications, as follows:
\begin{equation}%
\begin{array}
[c]{lll}%
&{\text{maximize} }&\sum_{l\in\mathcal{L}}U_l(r_l)
\\
&\text{subject to}& r_l=\min_{m\in\mathcal{M}_l}r_{lm}, \forall~l\in\mathcal{L}
\\
&& r_{lm}=\log(1+\gamma_{lm}(\mathbf{p})), \forall~l\in\mathcal{L}, m\in\mathcal{M}_l
\\
&& 0\leq p_l\leq P_l^{\max}, \forall l\in\mathcal{L}
\\
& \text{variables}& \{\mathbf{p},\{r_l\},\{r_{lm}\}\}.
\end{array}
\label{eq:sum_multicast}
\end{equation}

Similar to (\ref{eq:sum}), (\ref{eq:sum_multicast}) is nonconvex due to the complicated interference coupling between individual links. In order to devise distributed algorithms to solve (\ref{eq:sum_multicast}), it suffices to relax $r_l=\min_{m\in\mathcal{M}_l}r_{lm}$ in (\ref{eq:sum_multicast}) as $r_l\le\log(1+\gamma_{lm}(\mathbf{p})), \forall~l\in\mathcal{L}, m\in\mathcal{M}_l$. Thus, (\ref{eq:sum_multicast}) can be rewritten as
\begin{equation}%
\begin{array}
[c]{lll}%
&{\text{maximize} }&\sum_{l\in\mathcal{L}}U_l(r_l)
\\
&\text{subject to}& r_l\le\log(1+\gamma_{lm}(\mathbf{p})), \forall~l\in\mathcal{L}, m\in\mathcal{M}_l
\\
&& 0\leq p_l\leq P_l^{\max}, \forall l\in\mathcal{L}
\\
& \text{variables}& \{\mathbf{p},\mathbf{r}\}.
\end{array}
\label{eq:sum_multicast2}
\end{equation}

\subsection{Distributed Global Optimization Algorithms}

We develop next distributed algorithms that can find the globally optimal solutions to (\ref{eq:sum_multicast2}) based on EDT and SA. To this end, we first rewrite the optimization problem (\ref{eq:sum_multicast2}) as
\begin{equation}%
\begin{array}
[c]{lll}%
&{\text{minimize} }&-\sum_{l\in\mathcal{L}}U_l(r_l)
\\
&\text{subject to}& r_l\le\log(1+\gamma_{lm}(\mathbf{p})), \forall~l\in\mathcal{L}, m\in\mathcal{M}_l
\\
&& 0\leq p_l\leq P_l^{\max}, \forall l\in\mathcal{L}
\\
& \text{variables}& \{\mathbf{p},\mathbf{r}\}.
\end{array}
\label{eq:sum_multicast3}
\end{equation}

Next, we use EDT to write Lagrangian for (\ref{eq:sum_multicast3}) as
\begin{equation}%
\begin{array}
[c]{lll}%
L(\mathbf{p},\mathbf{r},\{\alpha_{lm}\})=-\sum_{l\in\mathcal{L}}U_l(r_l)\\
+\sum_{l\in\mathcal{L}, m\in\mathcal{M}_l}\alpha_{lm}(r_l-\log(1+\gamma_{lm}(\mathbf{p})))^{+},
\end{array}
\label{eq:penalty_multicast3}
\end{equation}
where $\alpha_{lm}\in \mathbb{R}$ are the penalty multipliers. Based on EDT, there exist finite $\alpha_{lm}^*\ge 0$ for all $l\in\mathcal{L}, m\in\mathcal{M}_l$ such that, for any $\alpha_{lm}>\alpha_{lm}^*$, $\forall~l\in\mathcal{L}, m\in\mathcal{M}_l$, the solution to (\ref{eq:penalty_multicast3}) is the same as (\ref{eq:sum_multicast3}) \cite{chen:2008}. Since there is no coupling among the power constraints of the individual users, (\ref{eq:penalty_multicast3}) does not include the power constraints. Thus, each user satisfies its own power constraint while minimizing (\ref{eq:penalty_multicast3}) in a distributed operation.

As in Section \ref{sec:pc_unicast}, the key step is to let each user perform a local stochastic search based on SA. Let $r_l$ and $p_l$ denote the primal values of the $l$th user, and $r_l'$ denote the new values randomly chosen by the $l$th user, which is treated as a new target transmission rate for the $l$th user. Different from the unicast communications case, the $l$th user updates $p_l'$ by
\begin{equation}%
\begin{array}
[c]{lll}%
p_l'=\min\left(\frac{e^{r_l'}-1}{\min_{m\in\mathcal{M}_l}\gamma_{lm}}p_l,P_l^{\max}\right),
\end{array}
\label{eq:pc_multicast}
\end{equation}
where $\gamma_{lm}$ is the current SINR measured at the receiver $m$ of user $l$. Note that (\ref{eq:pc_multicast}) does not need any information of the channel gains except the SINR feedback from the intended receivers, i.e.,  $\gamma_{lm}$.  Since (\ref{eq:pc_multicast}) is in standard form as described in \cite{yates:1995}, it converges geometrically fast to the target utility. The steps to update $r_l$ and $\alpha_{lm}$ are similar to DSPC Algorithm \ref{alg:dspc} in Section \ref{sec:pc_unicast}. A detailed description of DSPC algorithm for multicast communications is presented in Algorithm \ref{alg:dspc_multicast}.

\begin{prop}
\label{prop:dspc_multicast}
The distributed stochastic power control algorithm for multicast communications (Algorithm \ref{alg:dspc_multicast}) converges to a globally optimal solution to (\ref{eq:sum_multicast2}), as temperature $T$ in SA approaches zero.
\end{prop}
\begin{proof}
The proof is based on EDT and SA, and follows similar steps used in the proof of Proposition \ref{prop:dspc}, and it is omitted here for brevity.
\end{proof}

Likewise, to improve the convergence rate, we also propose an enhanced algorithm for Algorithm \ref{alg:dspc_multicast} by empirically choosing the initial penalty values and employing a geometric cooling schedule. The resulting algorithm is given in Algorithm \ref{alg:hdspc_multicast}. Similar to the unicast case, Algorithms \ref{alg:dspc_multicast} and \ref{alg:hdspc_multicast} do not need any knowledge of channel information (or the bottleneck link) and they are dynamically updated by the SINR feedback from the intended receivers.

\begin{algorithm}
\caption{DSPC for Multicast Communications}
\label{alg:dspc_multicast}
\begin{algorithmic}
\STATE \textbf{Initialization}: Choose $\epsilon>0$. Let $\alpha_{lm}=0$, $\forall~l\in\mathcal{L}, m\in\mathcal{M}_l$ and randomly choose $\mathbf{r}$ and $\mathbf{p}$.
\STATE \textbf{Step 1: update primal variables}
\STATE Set $T=T_0$, and select a sequence of time epochs $\{\tau_1,\tau_2,... \}$ in continuous time.
\STATE \quad \textbf{Repeat for each user} $l$
\begin{enumerate}
\STATE Randomly pick $r_l'$, and update $p_l'$ according to (\ref{eq:pc_multicast}).
\STATE Keep sensing the change of $\sum_{m\in\mathcal{M}_l}\alpha_{lm}(r_l-\log(1+\gamma_{lm}(\mathbf{p})))^{+}$ broadcast by other users.
\STATE Let $\Delta$ be the difference between $L(\mathbf{p},r_l|\mathbf{r}_{-l},\{\alpha_{lm}\})$ and $L(\mathbf{p}',r_l'|\mathbf{r}_{-l},\{\alpha_{lm}\})$, and accept $r_l'$ and $p_l'$ with probability 1, if $\Delta\ge 0$, or with probability $\exp(\frac{\Delta}{T})$, otherwise.
\STATE Broadcast $U_l(r_l')$, if $r_l'$ is accepted.
\STATE For each time epoch $\tau_i$, update $T=T_0/\log(i+1)$.
\end{enumerate}
\STATE \quad \textbf{Until} $T<\epsilon$.
\STATE \textbf{Step 2: update penalty variables}
\STATE \quad \textbf{For each user} $l$,
\begin{enumerate}
\STATE Update $\alpha_{lm}\leftarrow \alpha_{lm}+\varrho_{lm}(r_l-\log(1+\gamma_{lm}(\mathbf{p})))^{+}$, and scale down $\alpha_{lm}$, if the condition of penalty decrease is satisfied.
\STATE Goto Step 1 until no constraint is violated.
\end{enumerate}
\end{algorithmic}
\end{algorithm}

\begin{algorithm}
\caption{EDSPC for Multicast Communications}
\label{alg:hdspc_multicast}
\begin{algorithmic}
\STATE \textbf{Initialization}: Choose $\epsilon>0$. Let $\alpha_{lm}=\alpha_{lm}^0$, $\forall~l\in\mathcal{L}, m\in\mathcal{M}_l$ and randomly choose $\mathbf{r}$ and $\mathbf{p}$.
\STATE Set $T=T_0$, and select a sequence of time epochs $\{\tau_1,\tau_2,... \}$ in continuous time.
\STATE \textbf{Repeat for each user} $l$
\begin{enumerate}
\STATE Randomly pick $r_l'$, and update $p_l'$ according to (\ref{eq:pc_multicast}).
\STATE Keep sensing the change of $\sum_{m\in\mathcal{M}_l}\alpha_{lm}(r_l-\log(1+\gamma_{lm}(\mathbf{p})))^{+}$ broadcast by other users.
\STATE Let $\Delta$ be the difference between $L(\mathbf{p},r_l|\mathbf{r}_{-l},\{\alpha_{lm}\})$ and $L(\mathbf{p}',r_l'|\mathbf{r}_{-l},\{\alpha_{lm}\})$, and accept $r_l'$ and $p_l'$ with probability 1, if $\Delta\ge 0$, or with probability $\exp(\frac{\Delta}{T})$, otherwise.
\STATE Broadcast $U_l(r_l')$, if $r_l'$ is accepted.
\STATE For each time epoch $\tau_i$, update $T=\xi T$.
\end{enumerate}
\STATE \textbf{Until} $T<\epsilon$.
\end{algorithmic}
\end{algorithm}

\subsection{Numerical Examples}
In this section, we evaluate the performance of Algorithms \ref{alg:dspc_multicast} and \ref{alg:hdspc_multicast} for multicast communications. We consider a wireless network with four transmitters and each transmitter has two receivers.
These transmitters and receivers are randomly placed in a 10m-by-10m square area. The channel gains $h_{lm}$ are equal to $d_{lm}^{-4}$, where $d_{lm}$ represents the distance between the transmitter $l$ and the receiver $m$.
The channel gains $h_{lm}$ are equal to $d_{lm}^{-4}$, where $d_{lm}$ represents the distance between the transmitter $l$ and the receiver $m$. We assume $U_l(r_l)=r_l$, $P_l^{\max}=1$, and $n_{lm}=10^{-4}$ for all $l\in\mathcal{L}$ and $m\in\mathcal{M}_l$.
Fig. \ref{fig:convergence_multicast} illustrates the fast convergence performance of Algorithms \ref{alg:dspc_multicast} and \ref{alg:hdspc_multicast} in multicast communications.\footnote{The other existing algorithms have been specifically designed for unicast communications; therefore, they are excluded here from the performance comparison.} Besides, we examine the average performance (with confidence interval) of DSPC and EDSPC for multicast communications under 100 random initializations with the same system parameters as in Fig. \ref{fig:convergence_multicast}. As illustrated in Fig. \ref{fig:comparison_multicast}, both algorithms \ref{alg:dspc_multicast} and \ref{alg:hdspc_multicast} are robust against the initial value variations.

\begin{figure}[t]
\begin{center}
\vspace{-0.0cm}\hspace{-0.0cm} {\includegraphics[scale=0.5]{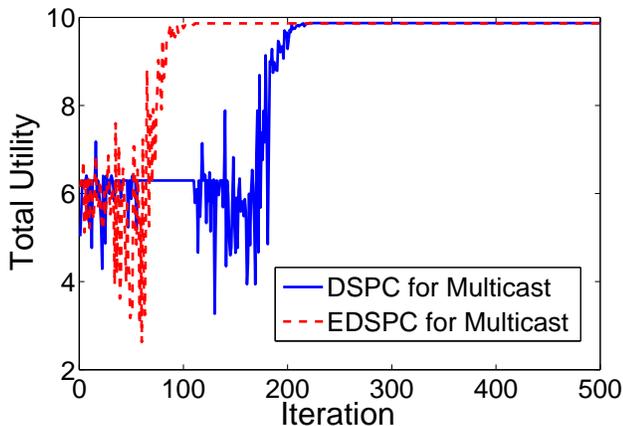}}\hspace{-0.0cm}
\caption{Convergence performance of DSPC and EDSPC for multicast communications.}
\label{fig:convergence_multicast}
\end{center}
\end{figure}



\begin{figure}[t]
\begin{center}
\vspace{0cm}\hspace{0cm} {\includegraphics[scale=0.5]{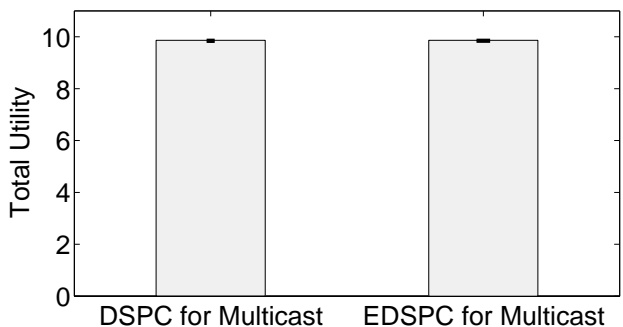}}\hspace{-0cm}
\vspace{-1.5cm} \caption{Comparison of average performance (with confidence interval) of DSPC and EDSPC for multicast.}\vspace{-0cm}
\label{fig:comparison_multicast}
\end{center}
\end{figure}

%

\section{Conclusion} \label{sec:conclusion}
We studied the distributed power control problem of optimizing the system utility as a function of the achievable rates in wireless ad hoc networks. Based on the observation that the global optimum lies on the boundary of the feasible region for unicast communications, we focused on the equivalent but more structured problem in the form of maximizing the minimum weighted utility. Appealing to extended duality theory, we decomposed the minimum weighted utility maximization problem into subproblems by using penalty multipliers for constraint violations. We then proposed a distributed stochastic power control (DSPC) algorithm to seek a globally optimal solution, where each user stochastically announces its target utility to improve the total system utility via simulated annealing. In spite of the nonconvexity of the underlying problem, the DSPC algorithm can guarantee global optimality, but only with a slow convergence rate. Therefore, we proposed an enhanced distributed algorithm (EDSPC) to improve the convergence rate with geometric cooling schedule in simulated annealing. We then compared DSPC and EDSPC with the existing power control algorithms and verified the optimality and complexity reduction.

Next, we proposed the joint scheduling and power allocation policy for queueing systems by integrating our distributed power control algorithms with the back-pressure algorithm. The stability region was evaluated, which is shown to be strictly greater than the throughput region in the corresponding saturated case. Beyond unicast communications, we generalized our power control algorithms to multicast communications by jointly maximizing the minimum rates on bottleneck links in different multicast groups. Our distributed stochastic power control approach guarantees global optimality  without the need of channel information, while reducing the computation complexity, in general systems with unicast and multicast communications, and applies to both backlogged and random traffic patterns.



\begin{thebibliography}{10}

\bibitem{yang:2011}
L. Yang, Y. E. Sagduyu, J. Zhang, and J. H. Li, ``Distributed power control for ad-hoc communications via stochastic nonconvex utility optimization,'' \emph{Proc. IEEE ICC}, 2011.

\bibitem{chiang:2008}
M. Chiang, P. Hande, T. Lan, and C. W. Tan, ``Power control in wireless cellular networks,'' \emph{Foundations and Trends in Networking}, vol. 2, no. 4, pp. 381-553, 2008.

\bibitem{julian:2002}
D. Julian, M. Chiang, D. O. Neill, and S. Boyd, ``Qos and fairness constrained convex optimization of resource allocation for wireless cellular and ad hoc networks,'' \emph{Proc. IEEE INFOCOM}, 2002.

\bibitem{chiang:2007}
M. Chiang, C. W. Tan, D. Palomar, D. O. Neill, and D. Julian, ``Power control by geometric programming,'' \emph{IEEE Trans. Wireless Commun.}, vol. 1, no. 7, pp. 2640-2651, 2007.

\bibitem{xiao:2003}
M. Xiao, N. B. Shroff, and E. K. P. Chong, ``A utility-based power control scheme in wireless cellular systems,'' \emph{IEEE/ACM Trans. Netw.}, vol. 11, no. 2, pp. 210-221, 2003.

\bibitem{luo:2008}
Z.-Q. Luo and S. Zhang, ``Dynamic spectrum management: complexity and duality,'' \emph{IEEE J. Sel. Topics Signal Process.}, vol. 2, no. 1, pp. 57-73, 2008.

\bibitem{huang:2006}
J. Huang, R. Berry, and M. Honig, ``Distributed interference compensation for wireless networks,'' \emph{IEEE J. Sel. Areas Commun.}, vol. 24, no. 5, pp. 1074-1084, 2006.

\bibitem{hande:2008}
P. Hande, S. Rangan, M. Chiang, and X. Wu, ``Distributed uplink power control for optimal SIR assignment in celluar data networks,'' \emph{IEEE/ACM Trans. Netw.}, vol. 16, no. 6, pp. 1420-1433, 2008.


\bibitem{saraydar:2002}
C. U. Saraydar, N. B. Mandayam, D. J. Goodman, ``Efficient power control via pricing in wireless data networks,'' \emph{IEEE Trans. Commun.}, vol. 50, no. 2, pp. 291-303, 2002.



\bibitem{alpcan:2002}
T. Alpcan, T. Basar, R. Srikant, and E. Altman, ``CDMA uplink power control as a noncooperative game,'' \emph{Wireless Networks}, vol. 8, no. 6, pp. 659-670, 2002.

\bibitem{qian:2009}
L. Qian, Y. J. Zhang, and J. W. Huang, ``MAPEL: achieving global optimality for a non-convex power control problem,'' \emph{IEEE Trans. Wireless Commun.}, vol. 8, no. 3, pp. 1553-1563, 2009.


\bibitem{qian:2010}
L. Qian, Y. J. Zhang, and M. Chiang, ``Globally optimal distributed power control for nonconcave utility maximization,'' \emph{Proc. IEEE GLOBECOM}, 2010.

\bibitem{geman:1984}
S. Geman and D. Geman, ``Stochastic relaxation, Gibbs distributions, and the Bayesian restoration of images,'' \emph{IEEE Trans. Pattern Anal. Mach. Intell.}, vol. 6, no. 6, pp. 721-741, 1984.


\bibitem{chen:2008}
Y. Chen and M. Chen, ``Extended duality for nonlinear programming,'' \emph{Comput. Optim. Appl.}, vol. 47, no. 1, pp. 33-59, 2010.

\bibitem{Kirkpatrick:1983}
S. Kirkpatrick, C. D. Gelatt, and J. M. P. Vecchi, ``Optimization by simulated annealing,'' \emph{Science}, vol. 220, no. 4598, pp. 671-680, 1983.


\bibitem{wieselthier:2000}
J. E. Wieselthier, G. D. Nguyen, and A. Ephremides, ``On construction of energy-efficient broadcast and multicast trees in wireless networks,''
\emph{Proc. IEEE INFOCOM}, 2000.

\bibitem{mo:2000}
J. Mo and J. Walrand, ``Fair end-to-end window-based congestion control,'' IEEE/ACM Trans. Netw., vol. 8, no. 5, pp. 556-567, 2000.

\bibitem{boyd:2004}
S. Boyd and L. Vandenberghe, \emph{Convex Optimization}. Cambridge, U.K.: Cambridge Univ. Press, 2004.



\bibitem{yates:1995}
R. D. Yates, ``A framework for uplink power control in cellular radio systems,'' \emph{IEEE J. Sel. Areas Commun.}, vol. 13, no. 7, pp. 1341-1347, 1995.




\bibitem{hajek:1988}
B. Hajek, ``Cooling schedules for optimal annealing,'' \emph{Math. Oper. Res.}, vol. 13, no. 2, pp. 311-329, 1988.

\bibitem{lin:2008}
X. Lin, N. B. Shroff, and R. Srikant, ``On the connection-level stability of congestion-controlled communication networks,'' \emph{IEEE Trans.
Inf. Theory}, vol. 54, no. 5, pp. 2317-2338, 2008.

\bibitem{ephremides:1992}
L. Tassiulas and A. Ephremides, ``Stability properties of constrained queueing systems and scheduling policies for maximum throughput in multihop radio networks,'' \emph{IEEE Trans. Autom. Control}, vol. 37, no. 12, pp. 1936-1948, 1992.

\bibitem{neely:2005}
M. J. Neely, E. Modiano, and C. E. Rohrs, ``Dynamic power allocation and routing for time varying wireless networks,'' \emph{IEEE J. Sel. Areas Commun.}, vol. 23, no. 1, pp. 89-103, 2005.

\bibitem{neely:2006}
L. Georgiadis, M. J. Neely, L. Tassiulas, ``Resource allocation and cross-layer control in wireless networks,'' \emph{Foundations and Trends in
Networking}, vol. 1, no. 1, pp. 1-149, 2006.

\bibitem{modiano:2011}
H.-W. Lee, E. Modiano, and L. B. Le, ``Distributed throughput maximization
in wireless networks via random power allocation,'' \emph{IEEE Trans. Mobile Comput.}, 2011.

\bibitem{sarkar:2008}
P. Chaporkar, S. Sarkar, ``Stable scheduling policies for maximizing throughput in generalized constrained queueing systems,'' \emph{IEEE Trans.  Autom. Control}, vol. 53, no. 8, pp. 1913-1931, 2008.

\bibitem{zhang:2009}
Y. Yi and J. Zhang and M. Chiang, ``Delay and effective throughput of wireless scheduling in heavy traffic regimes: vacation model for complexity,'' \emph{Proc. ACM Mobihoc}, 2009.

\bibitem{rao:1988}
R. Rao and A. Ephremides, ``On the stability of interacting queues in a multiple-access system,'' \emph{IEEE Trans. Inform. Theory}, vol. 34, no. 5, pp. 918-930, 1988.

\bibitem{parande:2008}
A. ParandehGheibi, M. Medard, A. Ozdaglar, A. Eryilmaz, ``Information theory vs. queueing theory for resource allocation in multiple access channels,'' \emph{Proc. IEEE Pers. Ind. Mob. Radio Commun.}, 2008.



\end{thebibliography}
\end{document}